\documentclass[11pt]{article}
\usepackage{axodraw}
\usepackage{epsfig}
\newcommand{\la}[1]{\label{#1}}
\newcommand{\be}{\begin{equation}}
\newcommand{\ee}{\end{equation}}
\newcommand{\ba}{\begin{eqnarray}}
\newcommand{\ea}{\end{eqnarray}}
\newcommand{\bi}{\begin{itemize}}
\newcommand{\ei}{\end{itemize}}
\newcommand{\rmi}[1]{{\mbox{\scriptsize #1}}}
\newcommand{\nr}[1]{(\ref{#1})}
\newcommand{\tr}{{\rm Tr\,}}

\newcommand{\Hc}{{\rm H.c.\ }}

\newcommand{\nn}{\nonumber \\}
\newcommand{\fr}[2]{{\frac{#1}{#2}}}

\newcommand{\bmu}{\bar\mu}
\newcommand{\1}{d} 
\newcommand{\2}{u} 
\newcommand{\3}{s} 
\newcommand{\4}{4} 
\renewcommand{\a}{r}    
\renewcommand{\b}{s}    
\renewcommand{\c}{u}    
\renewcommand{\d}{v}    
\renewcommand{\k}{k}    
\renewcommand{\l}{l}    
\newcommand{\ta}{\tilde r}     
\newcommand{\tb}{\tilde s}     
\newcommand{\tc}{\tilde u}     
\newcommand{\td}{\tilde v}     
\newcommand{\gvec}[1]{\bar{#1}}
\newcommand{\ga}{\gvec{\a}}
\newcommand{\gb}{\gvec{\b}}
\newcommand{\gc}{\gvec{\c}}
\newcommand{\gd}{\gvec{\d}}
\newcommand{\gk}{\gvec{\k}}
\newcommand{\gl}{\gvec{\l}}
\newcommand{\ua}{i}
\newcommand{\ub}{j}
\newcommand{\uc}{k}
\newcommand{\uk}{l}
\newcommand{\ia}{r}  
\newcommand{\ib}{u}  
\newcommand{\ic}{s}  
\newcommand{\id}{v}  
\newcommand{\iga}{\gvec{\ia}}
\newcommand{\igb}{\gvec{\ib}}
\newcommand{\igc}{\gvec{\ic}}
\newcommand{\igd}{\gvec{\id}}
\newcommand{\RR}{{\rm I\kern -.2em  R}}
\newcommand{\eq}{Eq.~}
\newcommand{\eqs}{Eqs.~}
\newcommand{\fig}{Fig.~}

\newcommand{\se}{Sec.~}

\newcommand{\Nf}{N_{\rm f}}
\newcommand{\Nc}{N_{\rm c}}

\def\lsi{\raise0.3ex\hbox{$<$\kern-0.75em\raise-1.1ex\hbox{$\sim$}}}
\def\gsi{\raise0.3ex\hbox{$>$\kern-0.75em\raise-1.1ex\hbox{$\sim$}}}

\makeatletter \@addtoreset{equation}{section} \makeatother
\renewcommand{\theequation}{\arabic{section}.\arabic{equation}}
\makeatletter
\renewcommand\section{\@startsection {section}{1}{\z@}%
                                   {-5.5ex \@plus -1ex \@minus -.2ex}
                                   {2.3ex \@plus.2ex}%
                                   {\normalfont\large\bfseries}}
\renewcommand\subsection{\@startsection{subsection}{2}{\z@}%
                                     {-3.25ex\@plus -1ex \@minus -.2ex}%
                                     {1.5ex \@plus .2ex}%
                                     {\normalfont\normalsize\bfseries}}
\renewcommand\thesection {\@arabic\c@section}
\renewcommand\thesubsection   {\thesection.\@arabic\c@subsection}
\renewcommand{\@seccntformat}[1]{%
\csname the#1\endcsname.\hspace{1.0em}}
\makeatother

\newcommand{\picc}[1]{\;\parbox[c]{60pt}{\begin{picture}(60,30)(0,0)
\SetWidth{1.0}\SetScale{1.0} #1 \end{picture}}\;}
\newcommand{\piccb}[1]{\;\parbox[c]{75pt}{\begin{picture}(75,30)(0,0)
\SetWidth{1.0}\SetScale{1.0} #1 \end{picture}}\;}
\newcommand{\piccc}[1]{\;\parbox[c]{90pt}{\begin{picture}(90,30)(0,0)
\SetWidth{1.0}\SetScale{1.0} #1 \end{picture}}\;}

\def\Asc(#1,#2)(#3,#4,#5){\CArc(#1,#2)(#3,#4,#5)}
\def\Lsc(#1,#2)(#3,#4){\Line(#1,#2)(#3,#4)}
\def\DAsc(#1,#2)(#3,#4,#5){\DashCArc(#1,#2)(#3,#4,#5){3}}
\def\DLsc(#1,#2)(#3,#4){\DashLine(#1,#2)(#3,#4){3}}
\def\TAsc(#1,#2)(#3,#4,#5){\SetWidth{2.0}\CArc(#1,#2)(#3,#4,#5)\SetWidth{1.0}}
\def\TLsc(#1,#2)(#3,#4){\SetWidth{2.0}\Line(#1,#2)(#3,#4)\SetWidth{1.0}}
\def\TDAsc(#1,#2)(#3,#4,#5){\SetWidth{2.0}\DashCArc(#1,#2)(#3,#4,#5){3}%
\SetWidth{1.0}}
\def\TDLsc(#1,#2)(#3,#4){\SetWidth{2.0}\DashLine(#1,#2)(#3,#4){3}%
\SetWidth{1.0}}

\def\CTopomeas(#1,#2){\picc{#1(0,15)(20,15) #2(20,15)(40,15)%
\SetWidth{2.0} \Line(17,12)(23,18) \Line(17,18)(23,12) \SetWidth{1.0}%
\GBoxc(0,15)(5,5){1} \GBoxc(40,15)(5,5){1} }}
\def\CTopomass(#1,#2){\picc{#1(0,15)(20,15) #2(20,15)(40,15)%
\GCirc(20,15){3}{0}%
\GBoxc(0,15)(5,5){1} \GBoxc(40,15)(5,5){1} }}
\def\CTopoL4(#1,#2){\picc{#1(0,15)(20,15) #2(20,15)(40,15)%
\GBoxc(20,15)(5,5){0}%
\GBoxc(0,15)(5,5){1} \GBoxc(40,15)(5,5){1}%
\Text(20,25)[c]{$L_4$} }}
\def\CTopoin(#1,#2,#3){\picc{#1(0,15)(20,15) #2(20,15)(40,15)%
#3(20,22)(7,0,360)%
\GBoxc(0,15)(5,5){1} \GBoxc(40,15)(5,5){1} }}
\def\CTopomassin(#1,#2,#3){\picc{#1(0,15)(20,15) #2(20,15)(40,15)%
#3(20,22)(7,0,360)%
\GCirc(20,15){3}{0}%
\GBoxc(0,15)(5,5){1} \GBoxc(40,15)(5,5){1} }}
\def\CTopocucu(#1,#2){\picc{#1(20,-5)(28,45,135) #2(20,35)(28,225,315)%
\GBoxc(0,15)(5,5){1} \GBoxc(40,15)(5,5){1} }}
\def\CTopocu(#1,#2){\picc{#1(0,22)(7,0,360) #2(0,15)(40,15)%
\GBoxc(0,15)(5,5){1} \GBoxc(40,15)(5,5){1} }}
\def\Topomeas(#1,#2,#3){\piccc{#1(0,15)(20,15) #2(20,15)(40,15)%
#3(40,15)(80,15) %
\SetWidth{2.0} \Line(17,12)(23,18) \Line(17,18)(23,12) \SetWidth{1.0}%
\GBoxc(0,15)(5,5){1} \GBoxc(80,15)(5,5){1} \GCirc(40,15){3}{1} }}
\def\Topomass(#1,#2,#3){\piccc{#1(0,15)(20,15) #2(20,15)(40,15)%
#3(40,15)(80,15) \GCirc(20,15){3}{0}%
\GBoxc(0,15)(5,5){1} \GBoxc(80,15)(5,5){1} \GCirc(40,15){3}{1} }}
\def\TopoL4(#1,#2,#3){\piccc{#1(0,15)(20,15) #2(20,15)(40,15)%
#3(40,15)(80,15) \GBoxc(20,15)(5,5){0}%
\GBoxc(0,15)(5,5){1} \GBoxc(80,15)(5,5){1} \GCirc(40,15){3}{1}%
\Text(20,25)[c]{$L_4$} }}
\def\Topoin(#1,#2,#3,#4){\piccc{#1(0,15)(20,15) #2(20,15)(40,15)%
#3(40,15)(80,15) #4(20,22)(7,0,360)%
\GBoxc(0,15)(5,5){1} \GBoxc(80,15)(5,5){1} \GCirc(40,15){3}{1} }}
\def\Topomassin(#1,#2,#3,#4){\piccc{#1(0,15)(20,15) #2(20,15)(40,15)%
#3(40,15)(80,15) #4(20,22)(7,0,360) \GCirc(20,15){3}{0}%
\GBoxc(0,15)(5,5){1} \GBoxc(80,15)(5,5){1} \GCirc(40,15){3}{1} }}
\def\Topoinop(#1,#2,#3){\piccc{#1(0,15)(38,8) #2(42,8)(80,15)%
#3(40,15)(7,0,360)%
\GBoxc(0,15)(5,5){1} \GBoxc(80,15)(5,5){1} \GCirc(40,22){3}{1} }}
\def\Topomassinop(#1,#2,#3){\piccc{#1(0,15)(38,8) #2(42,8)(80,15)%
#3(40,15)(7,0,360) \GCirc(40,8){3}{0}%
\GBoxc(0,15)(5,5){1} \GBoxc(80,15)(5,5){1} \GCirc(40,22){3}{1} }}
\def\Topocuop(#1,#2,#3){\piccc{#1(20,-5)(28,45,135) #2(20,35)(28,225,315)%
#3(40,15)(80,15)%
\GBoxc(0,15)(5,5){1} \GBoxc(80,15)(5,5){1} \GCirc(40,15){3}{1}}}
\def\Topocucu(#1,#2,#3){\piccc{#1(40,-25)(56,45,135) #2(0,15)(40,15)%
#3(40,15)(80,15)%
\GBoxc(0,15)(5,5){1} \GBoxc(80,15)(5,5){1} \GCirc(40,15){3}{1}}}
\def\Topocu(#1,#2,#3){\piccc{#1(0,22)(7,0,360) #2(0,15)(40,15)%
#3(40,15)(80,15)%
\GBoxc(0,15)(5,5){1} \GBoxc(80,15)(5,5){1} \GCirc(40,15){3}{1}}}
\def\Topocucuop(#1,#2,#3){\piccc{#1(20,-5)(28,45,135) #2(20,35)(28,225,315)%
#3(40,-25)(56,45,135)%
\GBoxc(0,15)(5,5){1} \GBoxc(80,15)(5,5){1} \GCirc(40,15){3}{1}}}
\def\Topoop(#1,#2,#3){\piccc{#2(0,15)(40,15) #3(40,15)(80,15)%
#1(40,22)(7,0,360)%
\GBoxc(0,15)(5,5){1} \GBoxc(80,15)(5,5){1} \GCirc(40,15){3}{1}}}
\def\Defmeas(#1,#2){\piccb{#1(0,15)(15,15) #2(15,15)(30,15)%
\SetWidth{2.0} \Line(12,12)(18,18) \Line(12,18)(18,12) \SetWidth{1.0}
\Text(30,15)[l]{ $=$ measure,} }}
\def\Defmass(#1,#2){\piccb{#1(0,15)(15,15) #2(15,15)(30,15)%
\GCirc(15,15){3}{0} \Text(30,15)[l]{ $=$ mass} }}
\def\Defmassin(#1,#2,#3,#4){\piccb{#1(0,15)(15,15) #2(15,15)(30,15)%
#3(15,15)(15,25) #4(15,15)(15,5)%
\GCirc(15,15){3}{0} \Text(30,15)[l]{ $=$ mass} }}
\def\Defcurr(#1){\picc{#1(0,15)(15,15) \GBoxc(0,15)(5,5){1} 
\Text(15,15)[l]{ $=$ current,} }}
\def\Defoper(#1,#2){\piccb{#1(0,15)(15,15) #2(15,15)(30,15)%
\GCirc(15,15){3}{1} \Text(30,15)[l]{ $=$ operator,} }}
\def\Deffield(#1,#2){\piccb{#1(0,15)(20,15)%
\Text(20,15)[l]{ $=$ #2,} }}
\begin{document}

\begin{titlepage}
\begin{flushright}
BI-TP 2004/15 \\
FTUV-04-0707 \\
IFIC/04-31 \\
hep-ph/0407086\\
\end{flushright}
\begin{centering}
\vfill

\mbox{\Large\bf Charm mass dependence of the weak Hamiltonian}

\vspace*{0.1cm}

\mbox{\Large\bf in chiral perturbation theory}

\vspace*{0.8cm}

P.~Hern\'andez$^{\rm a,}$\footnote{pilar.hernandez@ific.uv.es}
and 
M.~Laine$^{\rm b,}$\footnote{laine@physik.uni-bielefeld.de}

\vspace*{0.8cm}

{\em $^{\rm a}$%
Dpto.\ F\'{\i}sica Te\'orica and IFIC, Edificio Institutos Investigaci\'on, \\
Apt.\ 22085, E-46071 Valencia, Spain\\}

\vspace{0.3cm}

{\em $^{\rm b}$%
Faculty of Physics, University of Bielefeld, 
D-33501 Bielefeld, Germany\\}

\vspace*{0.8cm}

{\bf Abstract}
 
\end{centering}
 
\vspace*{0.4cm}

\noindent
Suppose that the weak interaction Hamiltonian of four-flavour SU(4) 
chiral effective theory is known, for a small charm quark mass $m_c$. 
We study how the weak Hamiltonian changes as the charm quark mass increases, 
by integrating it out within chiral perturbation theory to obtain 
a three-flavour SU(3) chiral theory. We find that the ratio of 
the SU(3) low-energy constants which mediate
$\Delta I=1/2$ and $\Delta I=3/2$ transitions, increases
rather rapidly with $m_c$, as $\sim m_c \ln (1/m_c)$.
The logarithmic effect originates from 
``penguin-type'' charm loops, and could represent one of the 
reasons for the $\Delta I=1/2$ rule.

\vfill



\vspace*{1cm}
 
\noindent
August 2004 

\vfill

\end{titlepage}

\section{Introduction}

The important observables of kaon physics, such as those related 
to CP violation, are sensitive to strongly interacting QCD dynamics. 
It has therefore remained a long-standing challenge to determine 
them reliably, starting from the Standard Model. 
The particular observable we concentrate on here is the 
so-called $\Delta I = 1/2$ rule, observed in non-leptonic 
strangeness violating weak decays, $K\to \pi\pi$
(for reviews see, e.g., Refs.~\cite{hg,revs}). 

There are a number of different physics scales relevant for these weak decays 
that could be responsible for the peculiar enhancement of the $\Delta I =1/2$ 
transitions. As the electroweak scale is decoupled through an operator
product expansion in the inverse W boson mass, large QCD radiative 
corrections could arise~\cite{wilson,itep}, but 
also physics at the charm quark 
mass scale of about a GeV~\cite{itep} or at the ``genuine'' QCD scale 
of a few hundred MeV~\cite{lattold} could be involved. 
Finally final state interactions 
at the pionic scale of about a hundred MeV have also been 
discussed as a possible source of the 
enhancement~\cite{lo}--\cite{pp}.

Given that the enhancement is a large one, it is important to establish 
whether its origin lies in the dynamics of one of these 
physics scales, or is the result of a fortuitous addition of several small 
effects. The best strategy to do this is to factorise as much as possible
the contributions from the different scales, and inspect them one at a time. 
While QCD corrections from the electroweak scale down can be 
addressed to a large extent within perturbation theory~\cite{wilson,itep}, 
the contribution of the QCD scale can be isolated 
particularly cleanly 
with lattice methods, by considering a world with an unphysically 
light charm quark, as has been recently 
proposed in Ref.~\cite{lattice}. Final state 
interactions, on the other hand, could be estimated within 
chiral perturbation theory~\cite{lo}--\cite{pp}.

Assuming that the programme 
of Ref.~\cite{lattice} has been carried out to completion 
and the weak Hamiltonian for the four-flavour SU(4) theory 
is known, the purpose of the present paper is to get 
a first impression of the effects of the charm quark mass, $m_c$, 
by studying how the $\Delta I=1/2$ and $\Delta I=3/2$ amplitudes evolve
as $m_c$ departs from the degenerate
chiral limit, but stays low enough so that 
chiral perturbation theory remains applicable to the charmed mesons. 
As the charm quark becomes heavy in comparison with the light quarks, 
the kaon weak decays can be described by a simpler effective theory,  
with three flavours only, the charm quark having been integrated out. 
If the charm is not too heavy, this SU(3) effective theory can be 
determined as a function of $m_c$ by matching it to the original 
SU(4) theory, employing chiral perturbation theory. The impact of 
$m_c$ on the $\Delta I=1/2$ rule has of course been widely acknowledged
a long time ago in the framework of the weak coupling
expansion~\cite{itep} and the large-$\Nc$ approach
(\cite{bbg,hpr} and references therein) but, to our knowledge, 
it has not been studied in this precise way before.

Obviously when the charm mass becomes large compared with the QCD scale,
as in the real world, charmed mesons become too heavy to be described
within the chiral theory. 
The determination of the couplings of the SU(3) effective 
theory requires in this situation lattice methods~\cite{b}.

The plan of this paper is the following. In~\se\ref{se:weak}
we briefly recall the form of the weak interaction Hamiltonian
in SU(4)$_L \times$SU(4)$_R$ chiral effective theory. 
In~\se\ref{mc} we set up the formalism for integrating out
the charm quark. The computation of the relevant matching 
coefficients is presented in~\se\ref{se:matching}. We discuss
the physical implications of our results in~\se\ref{se:physics}, 
and conclude in~\se\ref{se:concl}. 

\section{Weak effective Hamiltonian in SU(4) chiral perturbation theory}
\la{se:weak}

For energy scales below a few hundred MeV, the physics of QCD 
can be described by an effective chiral theory. Considering the unphysical
situation of a light charm quark and ignoring weak interactions, 
the chiral theory possesses an SU(4)$_L\times$SU(4)$_R$ symmetry, 
broken ``softly'' by the mass terms. The Euclidean Lagrangian 
can to leading order be written as 
\be
 \mathcal{L}_E = 
 \frac{F_{}^2}{4} \tr \Bigl[\partial_\mu U\partial_\mu U^\dagger \Bigr] - 
 \frac{\Sigma}{2} \tr \Bigl[ 
 U M                   
 + M^\dagger U^\dagger 
 \Bigr]. 
 \la{LE}
\ee
Here $U\in $ SU(4), 
$M$ is the quark mass matrix and, 
to leading order in the chiral expansion,  $F_{}$, $\Sigma$, 
equal the pseudoscalar decay constant and 
the chiral condensate, respectively. 
We will for convenience take $M$ to be real and diagonal, and denote 
\be
 \chi = \frac{2 \Sigma M^\dagger}{F_{}^2};  
\ee
for degenerate quark masses,  
the pion mass is then (to leading order) $M_\pi^2 = {[\chi]}_{11}$. 

Since the theory in \eq\nr{LE} is non-renormalisable, 
higher order operators usually contribute at next-to-leading order. 
We will here need explicitly only operators of the form~\cite{gl2}
\be
 \delta \mathcal{L}_E = 
 L_4 \, \tr [\partial_\mu U\partial_\mu U^\dagger] \,
 \tr[\chi^\dagger U + U^\dagger \chi] 
 - L_6 \Bigl( \tr[\chi^\dagger U + U^\dagger \chi] \Bigr)^2. 
 \la{higher_order}
\ee
There are phenomenological estimates available for the numerical values 
of $L_4,L_6$, in the physical SU(3)$_L\times$SU(3)$_R$ 
symmetric case~\cite{gl2,beg}.

Weak interactions break explicitly
the SU(4)$_L\times$SU(4)$_R$ symmetry of \eq\nr{LE}. 
In the chiral theory, the corresponding generic operator can 
to leading order be written as~\cite{hg,revs}
\ba
 {[{\cal O}_{w}]}_{\a\b\c\d} & \equiv &  
 \fr14 F_{}^4 
 \Bigl(\partial_\mu U U^\dagger\Bigr)_{\c\a}
 \Bigl(\partial_\mu U U^\dagger\Bigr)_{\d\b}\;. 
 \la{O_XPT} 
\ea
This operator is singlet under SU(4)$_R$, while
the reduction to irreducible representations of SU(4)$_L$
(denoted by $[\hat {\cal O}_{w}]^\sigma_{\a\b\c\d}$, etc) 
is summarised in~\ref{app:su4}. Of course, there 
are other operators with the same symmetries, but 
a higher order in the chiral expansion: these could either
involve more derivatives (a complete collection of such 
next-to-leading order operators can be found in Ref.~\cite{p4}),
or explicit occurrences of the mass matrix, as in 
\ba
 {[{\cal O}_m]}_{\a\b\c\d} & \equiv & -\frac{\Sigma}{2} \Bigl\{ 
  (M^\dagger M)_{\c\b} 
  (U M 
  + M^\dagger U^\dagger)_{\d\a} 
  + (M^\dagger M)_{\d\a} 
  (U M + M^\dagger U^\dagger)_{\c\b}    
  \Bigr\}
  \;.\la{O2_XPT}
\ea
For reasons to become clear later on it is sufficient,
however, to concentrate here on ${\cal O}_{w}$. 

In the CP conserving case of two generations, the 
strangeness violating part of 
the chiral weak Hamiltonian can then be written as
\be
 {\cal H}_w =  
 \sum_{\sigma=\pm 1} 
 g_{w}^\sigma {\hat c}_{\a\b\c\d}^\sigma
 [\hat {\cal O}_{w}]_{\a\b\c\d}^\sigma 
 + ... + \Hc \; . 
 \la{Lw_XPT}
\ee
Here $g_{w}^\sigma$ are dimensionful constants,  
${\hat c}^\sigma_{\a\b\c\d}$ are Clebsch-Gordan type pure numbers,  
and $\sigma = \pm 1$ correspond to irreducible
representations of SU(4)$_L$ with dimensions 84, 20, respectively.

For later reference let us note that
if we choose to write ${\hat c}_{\a\b\c\d}^\sigma$ in an unsymmetrised 
form, ${\hat c}_{\a\b\c\d}^\sigma \to c_{\a\b\c\d}$, then $c_{\a\b\c\d}$ 
can be taken to be~\cite{hg}
\ba
 & & c_{suud} = 1,\quad c_{sccd} = -1\;. 
 \la{physind}
\ea
The other $c_{\a\b\c\d}$ can be assumed zero, although the 
symmetries of the operators $[\hat {\cal O}_{w}]_{\a\b\c\d}^\pm$ imply 
that effectively only properly symmetrised parts of $c_{\a\b\c\d}$ 
contribute (cf.~\eq\nr{sym_c}). 
We may also recall that the ``tree-level''
(or large-$\Nc$, since $\alpha_s \sim 1/\Nc \to 0$) values 
for $g_{w}^\pm$ are $g_{w}^\pm = 2 \sqrt{2} G_F V_\rmi{ud} V^*_\rmi{us}$,
where $G_F$ is the Fermi constant, and $V_\rmi{ud}, V_\rmi{us}$ 
are elements of the CKM matrix.
We will not use the tree-level values, though, but assume that 
$g_{w}^\pm$ have been determined non-perturbatively, for instance
with the procedure outlined in Ref.~\cite{lattice}. 


\section{Decoupling of the charm quark}
\label{mc}

We now take a mass matrix of the form 
\be
 M = \mathop{\mbox{diag}}(m_u,m_d,m_s,m_c),  \la{Mass}
\ee
with $m_c \gg m_u,m_d,m_s$. 
For simplicity we assume for the moment that
$m_u = m_d = m_s$, although this does not affect our actual results. Let 
us denote 
\be
 M_u^2 \equiv \frac{m_u \Sigma}{F_{}^2}, \qquad
 M_c^2 \equiv \frac{m_c \Sigma}{F_{}^2}, \qquad
 M_{uc}^2 \equiv M_u^2 + M_c^2.
\ee

If we consider momenta smaller than the mass scale $M_c^2$, 
then we expect the physics of the SU(4)$_L\times$SU(4)$_R$
theory in~\eq\nr{LE} to be contained in another theory 
from which the heavy scale has been integrated out, and which thus 
has  an SU(3)$_L\times$SU(3)$_R$ symmetry.  
We would like to derive the effective weak Hamiltonian of such
a theory, given some non-perturbative values for the 
coefficients $g_{w}^\pm$ in~\eq\nr{Lw_XPT}, and assuming
$M_u^2  \ll M_c^2 \ll (4 \pi F_{})^2$.  
The first of these hierarchies, $M_u^2 \ll M_c^2$, will allow us to 
truncate the effective action by dropping higher order operators 
suppressed parametrically
by $1/M_c^n$, with some power $n$.
The lowest-order non-singlet 
building blocks for weak operators in the SU(3) theory will be  
listed in~\eqs\nr{su3_1}, \nr{su3_3}  below. The second hierarchy, 
$M_c^2 \ll (4 \pi F_{})^2$,  is necessary
to ensure that the charmed mesons can be treated 
within SU(4)$_L\times$SU(4)$_R$ chiral
perturbation theory, and it makes its appearance in our expressions
for the coefficients of the operators that are kept; we work 
out the first two orders, 
i.e.,~${\cal O}(1)$ and ${\cal O}(M_c^2/(4 \pi F_{})^2)$, 
for the operators in~\eqs\nr{su3_1}, \nr{su3_3}.    

The form of the effective
SU(3)$_L\times$SU(3)$_R$ chiral theory is just like~\eq\nr{LE}, 
only with modified parameters, and SU(3) matrices. 
We will, in general, distinguish
the parameters and observables of the SU(3)$_L\times$SU(3)$_R$ theory
from those of the SU(4)$_L\times$SU(4)$_R$ theory with a bar:
\be
 \bar {\cal L}_E = 
 \frac{\bar F_{}^2}{4} \tr \Bigl[\partial_\mu \bar U\partial_\mu 
 \bar U^\dagger \Bigr] - 
 \frac{\bar \Sigma}{2} \tr \Bigl[ 
 \bar U \bar M                   
 + \bar M^\dagger \bar U^\dagger 
 \Bigr],
 \la{bLE}
\ee
where
\be
 \bar M = \mathop{\mbox{diag}}(m_u,m_d,m_s) \;. 
\ee
The SU(3) flavour indices are denoted by $\ga,\gb,\gc,\gd$ . 

Non-singlet weak operators can now be built with the SU(3) analogue of 
the operator in~\eq\nr{O_XPT}, 
but also with 
operators transforming under ${\bf 3^*} \otimes {\bf 3}$ of
SU(3)$_L$~\cite{b}.  Thus, we will encounter
\ba
 {[\bar {\cal O}_{w}]}_{\ga\gb\gc\gd} & \equiv &  
 \fr14 \bar F_{}^4 
 \Bigl(\partial_\mu \bar U \bar U^\dagger\Bigr)_{\gc\ga}
 \Bigl(\partial_\mu \bar U \bar U^\dagger\Bigr)_{\gd\gb}\;, 
 \la{su3_1} \\
 {[\bar {\cal O}_{m}]}_{\ga\gc} & \equiv & \fr12 \bar F_{}^2 {\bar \Sigma}
   (\bar U \bar M 
   + \bar M^\dagger \bar U^\dagger)_{\gc\ga} 
   \;.
 \la{su3_3}
\ea
The weak Hamiltonian is denoted by $\bar {\cal H}_w$, and our objective
is to find the coefficients with which ${[\bar {\cal O}_{w}]}_{\ga\gb\gc\gd}$, 
${[\bar {\cal O}_{m}]}_{\ga\gc}$ appear there, given ${\cal H}_w$ 
in~\eq\nr{Lw_XPT}.

The leading order SU(3) operators having been identified, the scale hierarchy
$M_u^2 \ll M_c^2$ has now been taken care of. The remaining 
challenge is then to compute the coefficients of these operators. 
The coefficients are in general some complicated functions of $m_c$
and the QCD scale, but in the chiral regime to which we restrict 
ourselves in this paper, the remaining hierarchy $M_c^2 \ll (4 \pi F_{})^2$ 
allows us to determine them explicitly to some finite order in a Taylor 
series in $M_c^2 / (4 \pi F_{})^2$, 
as we do in the next section. 


\section{Matching for the coefficients}
\la{se:matching}

\subsection{Basic setup}

In order to determine $\bar F_{}, \bar \Sigma$ 
as well as the coefficients of $\bar {\cal O}_{w}$, 
$\bar {\cal O}_{m}$  in $\bar {\cal H}_w$,
we match the predictions for various
non-vanishing observables computable both in the original and
in the effective theory.\footnote{%
   A completely analogous computation, namely finding
   relations between $\bar F_{}, \bar \Sigma$ and the corresponding
   parameters of an SU(2)$_L\times$SU(2)$_R$ chiral effective theory, 
   obtained after integrating out the strange quark, was carried out 
   already in Ref.~\cite{gl2} (for recent work and references see, 
   e.g., Ref.~\cite{sde}).
   } 
For this purpose, we define correlators 
involving left-handed flavour currents (this choice is well suited
also for a practical implementation of matching computations
on the lattice~\cite{algo,us}). 
Such correlators are computed 
by promoting the partial derivatives in~\eqs\nr{LE}, \nr{higher_order}, 
\nr{bLE} into covariant ones, 
\be
 \partial_\mu U \to D_\mu U \equiv [\partial_\mu + i A^a_\mu \bar T^a] U,
 \qquad
 \partial_\mu \bar U \to D_\mu \bar U \equiv 
 [\partial_\mu + i A^a_\mu \bar T^a] \bar U,
 \ee
where $\bar T^a$ are Hermitean generators of SU(3),\footnote{%
  Or, more precisely in the case of SU(4), Hermitean generators 
  in the subalgebra which generates SU(3).}
and then taking a second functional derivative
with respect to these fields~\cite{gl2}. However, apart from 
``contact'' contributions (arising from operators overlapping at the 
same spacetime location) and ``counterterm'' contributions 
(arising from~\eq\nr{higher_order}), we can as well write down the
correlators directly in terms of the left-handed currents,    
\be
 {\cal J}_\mu^a \equiv
 \left. \Bigl( \frac{\partial {\cal L}_E}{\partial A^a_\mu} \Bigr)
 \right|_{A^a_\mu = 0} 
 = -i \frac{F_{}^2}{2} \bar T^a _{\ia\ib} \Bigl(
 \partial_\mu U U^\dagger \Bigr)_{\ib\ia} + ... \; ,
 \la{cJmua}
\ee
and correspondingly for $\bar {\cal J}^a_\mu$.

To leading order in the weak 
Hamiltonian, matching can thus be carried out by requiring
\ba
 \Bigl\langle {\cal J}^a_\mu(x) {\cal J}^b_\nu(y) 
 \Bigl\rangle_\rmi{SU(4)} & = &  
 \Bigl\langle \bar {\cal J}^a_\mu(x) \bar {\cal J}^b_\nu(y) 
 \Bigl\rangle_\rmi{SU(3)}, \la{bm1} \\
 \Bigl\langle {\cal J}^a_\mu(x) 
 \; {\cal H}_w(z) \;  
 {\cal J}^b_\nu(y) 
 \Bigl\rangle_\rmi{SU(4)} & = &  
 \Bigl\langle \bar {\cal J}^a_\mu(x) 
 \; \bar {\cal H}_w(z) \;  
 \bar {\cal J}^b_\nu(y) 
 \Bigl\rangle_\rmi{SU(3)}, \la{match_su3}
\ea
where the expectation values are evaluated using 
the strangeness conserving Lagrangian, and space-time 
separations between the sources and the weak Hamiltonian
are assumed large compared with $M_c^{-1}$.

Since the result of the matching computation is insensitive to 
infrared physics we will, for simplicity, carry out the 
computation in an infinite volume.  
We may then write 
\be
 U = \exp\Bigl({i \frac{2\xi}{F_{}}}\Bigr),  \la{U}
\ee
where $\xi$ is traceless and Hermitean. 
For the SU(4) indices,
we introduce the somewhat implicit shorthand notation that
\be
 f_{\ga...} \equiv 
 \Bigl(\delta_{\a\b}-\delta_{\a \4}\delta_{\b \4} \Bigr) f_{\b...}.
 \la{def_barred}
\ee 
Here the index 4 refers to the charm flavour that is to be integrated out.
Thus, $f_{\ga...}$ is non-trivial only for index values $\a = u,d,s$,
and we can write a general SU(4) tensor as
\ba
 f_{\a...} = f_{\ga...} +  \delta_{\a \4} f_{\4...}.  
\ea
Consequently, expanding~\eq\nr{U} inside~\eq\nr{LE}, 
the SU(4) free propagator becomes
\ba
 \Bigl\langle 
 \xi_{\ib\ia}(x)\xi_{\id\ic}(0) \Bigr\rangle & = &  
 \fr12 \Bigl( \delta_{\igb\igc} \delta_{\iga\igd} - 
       \fr13 \delta_{\igb\iga} \delta_{\igd\igc} \Bigr)
 G(x;2 M_u^2) \nn 
 & + &     
 \fr12 (\delta_{\igb\igc} \delta_{\ia \4} \delta_{\id \4} + 
 \delta_{\iga\igd}\delta_{\ib \4}\delta_{\ic \4}
 )\, G(x; M_{uc}^2) \nn 
 & + &  
 \fr1{24}(\delta_{\igb\iga} - 3 \delta_{\ib \4}\delta_{\ia \4})
         (\delta_{\igd\igc} - 3 \delta_{\id \4}\delta_{\ic \4})\, 
 G(x;M_u^2/2 + 3M_c^2/2), 
 \la{propag}
\ea
where 
\be
 G(x;M^2) = \int \! \frac{{\rm d}^{d} p}{(2\pi)^{d}} 
 \frac{e^{i p\cdot x}}{p^2 + M^2}
 \;, \la{Gx}
\ee
and $d$ is the dimension of the spacetime. 
At the order of the chiral expansion we are working, most
non-trivial effects come from the second term 
on the right-hand-side of~\eq\nr{propag}.

On the SU(3) side, we write 
\be
 \bar U = \exp\Bigl({i \frac{2\bar \xi}{\bar F_{}}}\Bigr),  \la{barU}
\ee
and the free propagator is
\be
 \Bigl\langle \bar \xi_{\igb\iga}(x) \bar \xi_{\igd\igc}(0) \Bigr\rangle =
 \fr12 \Bigl(\delta_{\igb\igc} \delta_{\iga\igd}  - 
          \fr13 \delta_{\igb\iga} \delta_{\igd\igc} 
 \Bigr)\, G(x;2 \bar M_u^2), \la{def_xixi}
\ee 
where $\bar M_u^2 = m_u \bar \Sigma/ \bar F_{}^2$.

The objective, then, is to compute the observables 
on the left-hand-sides of~\eqs\nr{bm1}, \nr{match_su3} 
to order ${\cal O}(M_c^2/(4\pi F)^2)$, and find out the  
modified parameters $\bar F, \bar \Sigma$ and the coefficients
of the non-singlet weak operators in~\eqs\nr{su3_1}, \nr{su3_3}
that reproduce the same results, 
in terms of the couplings of the SU(4) theory and $m_c$.

\subsection{Pion decay constant and chiral condensate}
\la{match_F0}

Consider first a correlator of the form in~\eq\nr{bm1}. 
On the effective SU(3) theory side, we only 
need the lowest order result for this correlator,
\be
 \Bigl\langle \bar {\cal J}^a_\mu(x) 
 \bar {\cal J}^b_\nu(y) \Bigr\rangle = 
 - \bar F_{}^2 \; \bar T^a_{\iga\igb} \bar T^b_{\igc\igd} \;
 \partial_\mu^x \partial_\nu^x \,
 \Bigl\langle \bar \xi_{\igb\iga}(x) \bar \xi_{\igd\igc}(y) \Bigr\rangle, 
 \la{treeC}
\ee
where the propagator is in~\eq\nr{def_xixi}. 

Consider then the corresponding observable on the SU(4) side, 
now at 1-loop level. Let us inspect its behaviour
at large separations, $|x-y| \gg M_c^{-1}$. 
This means that any graph 
including other than tadpole type 
internal lines of the heavy field cannot contribute, 
since the result is exponentially suppressed. 
On the other hand, there are non-trivial 1-loop effects 
from the graphs in~\fig\ref{fig:effF} that do contribute.

\begin{figure}[t]
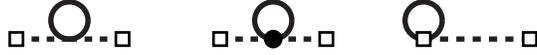


\begin{eqnarray*}
& & 
\CTopoin(\TDLsc,\TDLsc,\TAsc)  \quad
\CTopomassin(\TDLsc,\TDLsc,\TAsc) \quad
\CTopocu(\TAsc,\TDLsc) 
\end{eqnarray*}

\caption[a]{\small 
The graphs computed in \se\ref{match_F0}. 
Dashed and solid lines denote light and heavy fields, respectively, 
and an open box is the left-handed
current. A ``bare'' (four-point) vertex and 
a filled circle originate from the first and second 
terms in~\eq\nr{LE}, respectively.}

\la{fig:effF}
\end{figure}

The result for the graphs in~\fig\ref{fig:effF}, together
with the contribution from the counterterm in~\eq\nr{higher_order},  
can be written (after carrying out a partial integration, 
and taking the limit $M_u^2/M_c^2 \to 0$) in the form
\be
 \Bigl\langle {\cal J}^a_\mu(x) 
 {\cal J}^b_\nu(y) \Bigr\rangle =
 - F_{}^2 \; \bar T^a_{\iga\igb} \bar T^b_{\igc\igd}\; 
 \partial_\mu^x \partial_\nu^x \;
 \Bigl\langle \xi_{\igb\iga}(x) \xi_{\igd\igc}(y) \Bigr\rangle
 \Bigl(1 -  
 \frac{G(0;M_c^2)}{F_{}^2}  + 
 16 L_4 \frac{M_c^2}{F_{}^2} \Bigr), 
 \la{JJsu4}
\ee
where the behaviour of
$\bar T^a_{\iga\igb} \bar T^b_{\igc\igd}\;
\langle \xi_{\igb\iga}(x) \xi_{\igd\igc}(y) \rangle$ 
is determined by the first term in~\eq\nr{propag}, 
with $M_u^2 \to 0$. 
Comparing with~\eq\nr{treeC}, we arrive at 
an expression for the SU(3) pion decay constant in the chiral limit,
\be
 \bar F_{}^2 = F_{}^2 \biggl\{ 
 1 - \frac{1}{F_{}^2} \biggl[ G(0;M_c^2) - 16 L_4 M_c^2 \biggr] 
 \biggr\}. \la{barF}
\ee
This is a finite correction, the logarithmic 
divergence of $G(0;M_c^2)$ being cancelled by 
the known one~\cite{gl2} in $L_4$. Note that the relation 
in \eq\nr{barF} is completely analogous to that 
between the pion decay constants of the SU(3) and 
SU(2) chiral effective theories~\cite{gl2}, the role
of $m_s$ just now played by $m_c$. 

It is also easy to compute the relation between $\bar\Sigma, \Sigma$.
This can be inferred either by keeping the term proportional to $M_u^2$ 
in~\eq\nr{JJsu4}, or directly by matching 
the predictions for the chiral condensate. Setting again $M_u^2/M_c^2 \to 0$
after the matching, we obtain the SU(3) parameter $\bar \Sigma$ in the chiral
limit at next-to-leading order in $M_c^2/(4\pi F)^2$, 
\be
 \bar \Sigma  = \Sigma \biggl\{
 1 - \frac{1}{F_{}^2}\biggl[
 G(0;M_c^2) + \fr1{12} G(0;3 M_c^2/2) - 32 L_6 M_c^2
 \biggr]
 \biggr\}\;. \la{Sigma}
\ee
The result cited for the divergent part of $L_6$ in Ref.~\cite{gl2} is 
specific for SU(3), but the propagators in~\eq\nr{propag} and result 
in~\eq\nr{Sigma} are easily generalised to any $\Nf$. This way 
we find that the divergence scales as\footnote{%
  Expressions for the divergences at 
  a general $\Nf$ can also be found in Ref.~\cite{cp}, for instance.} 
$1+2/\Nf^2$, and multiplying the
result of Ref.~\cite{gl2} by the corresponding ratio of flavour factors, the 
logarithmic divergence is cancelled by the one in~$L_6$. 
Again, the relation in~\eq\nr{Sigma} is completely analogous 
to the known relation between chiral condensates in the SU(3) and 
SU(2) chiral effective theories~\cite{gl2}.

\subsection{A generic weak operator ${\cal O}_{w}$}
\la{match_weak}

Next, we consider the weak part of the SU(4) theory,~\eq\nr{Lw_XPT}.
This leads to a corresponding weak part in the SU(3) theory. 
It is convenient to start by considering the matrix element of the 
SU(4) operator, $[{\cal O}_{w}]_{\a\b\c\d}$, 
with two left currents.\footnote{%
  That is, an observable of the type on the left-hand side 
  of~\eq\nr{match_su3} but with 
  ${\cal H}_w(z) \to [{\cal O}_{w}]_{\a\b\c\d}(z)$.} 
Given that the left currents 
have SU(3) indices (cf.\ \eq\nr{cJmua}), 
this matrix element should 
at large distances ($|x-z|, |y-z|, |x-y|  \gg M_c^{-1}$) 
be reproduced
in the SU(3) theory by the matrix element of a combination of the 
operators in~\eqs\nr{su3_1}, \nr{su3_3} between two SU(3) left currents. 
The result for the full matrix 
element of ${\cal H}_w$ in~\eq\nr{match_su3}  
can then be completed by summing over 
all operators with the 
correct SU(4) and SU(3) weights and 
classifications, as we will do in~\se\ref{su3_class}.

In order to carry out this matching, we actually 
do not need to compute the full 
three-point functions. First of all, we realize that 
the only diagrams with heavy propagators that are unsuppressed
 at large distances are those in~\fig\ref{fig:effc} 
(i.e.,\ those with closed heavy quark loops), 
since any diagram with heavy propagators
connecting different space-time points is 
exponentially suppressed at large distances. 
Among the graphs in~\fig\ref{fig:effc}, the ones on the first row involve
a weak operator with the index structure $[{\cal O}_{w}]_{\ga \gb \gc \gd}$.  
Their contribution can easily be  seen to be of the 
form $\Bigl\langle \bar {\cal J}^a_\mu(x) 
 \; [ \bar {\cal O}_w ]_{\ga \gb \gc \gd}(z) \; \bar {\cal J}^b_\nu(y)  
\Bigl\rangle_\rmi{SU(3)}$, up to a coefficient. 
Obviously this is the case also for the graphs involving 
only light propagators. 

\begin{figure}[t]
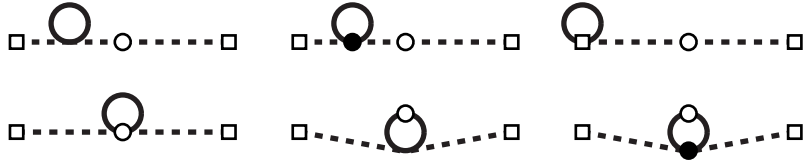


\begin{eqnarray*}
& &
\Topoin(\TDLsc,\TDLsc,\TDLsc,\TAsc) \quad
\Topomassin(\TDLsc,\TDLsc,\TDLsc,\TAsc) \quad
\Topocu(\TAsc,\TDLsc,\TDLsc) \\
& & 
\Topoop(\TAsc,\TDLsc,\TDLsc) \quad 
\Topoinop(\TDLsc,\TDLsc,\TAsc) \quad
\Topomassinop(\TDLsc,\TDLsc,\TAsc) 
\end{eqnarray*}

\caption[a]{\small 
The graphs computed in \se\ref{match_weak}. 
An open circle denotes a weak operator ${\cal O}_{w}$, otherwise the 
notation is as in \fig\ref{fig:effF}.}

\la{fig:effc}
\end{figure}

On the other hand, in the graphs on the second row 
in~\fig\ref{fig:effc} one has  
two indices ``\4'' in $[{\cal O}_{w}]_{\a\b\c\d}$, 
which are contracted in various ways. The 
matching of these contributions is a little bit more subtle 
so we will give a few details.  

Suppose the operator (e.g.\ $[{\cal O}_{w}]_{\ga 44\gd}$) is located
at the point $z$, and we close the loop in the last three graphs
of \fig\ref{fig:effc}, while keeping the light legs still as 
uncontracted fields ($\xi$). We denote such an integration 
over the heavy fields by $\langle ... \rangle_\4$.
It should be obvious that the final contribution 
can be obtained from the matrix elements 
$\langle {\cal J}^a_\mu(x) \; 
\langle {\cal O}(z) \rangle_\4 \;{\cal J}^b_\nu(y)\rangle$ after 
integrating over the light fields, so the matching 
can be accomplished
by identifying the structures $\bar {\cal O}_{w}$ and $\bar {\cal O}_{m}$ 
in the averages $\langle {\cal O}(z) \rangle_\4$. 
The result of this computation reads:
\ba
 & & \hspace*{-1.2cm} \langle [{\cal O}_{w}]_{\ga 44\gd}(z) \rangle_\4
 = 
   -\fr16 \xi_{\gd \gk}(z) \xi_{\gk \ga}(z)\;\partial_\omega^2 G(0;M_{uc}^2)-
  \fr12 \partial_\mu \xi_{\gd \gk}(z) \partial_\mu \xi_{\gk \ga}(z)
  \;G(0;M_{uc}^2)
 \nn  & & \hspace*{0.2cm}
 -\fr16 \int_s \xi_{\gd\gk}(s)\xi_{\gk\ga}(s) 
 \Bigl\{
 M_c^2 [\partial_\omega G(s-z;M_{uc}^2)]^2 
 + [\partial_\omega \partial_\psi G(s-z;M_{uc}^2)]^2 
 \Bigr\} \nn
 & & \hspace*{0.2cm} -\fr16 \int_s \Bigl\{  
 \partial_\mu \xi_{\gd\gk}(s) \partial_\mu \xi_{\gk\ga}(s)
 + \fr12 (\partial_\mu^2 + 6 M_u^2) 
 [ \xi_{\gd\gk}(s)\xi_{\gk\ga}(s) ]
 \Bigr\} [\partial_\omega G(s-z;M_{uc}^2)]^2, 
 \la{expl}
\ea
where $s$ is the location of the four-point interaction, 
and $\int_s \equiv \int {\rm d}^ds$. 

In order to illustrate the structures
appearing after the remaining integration
in a transparent way, 
we shall for the moment leave out all indices 
and derivatives, and rewrite~\eq\nr{expl} symbolically as
\be
 \langle {\cal O}(z) \rangle_\4 = 
 \int_s \xi(s) \xi(s) g(s-z;M_{uc}^2)
 \;,  \la{mat1}
\ee
where $g(s-z;M_{uc}^2)$ 
is a function which is
exponentially suppressed for $|s-z| \gg M_c^{-1}$. 
In order to expand~\eq\nr{mat1} in terms of local operators, 
we go to momentum space, i.e.\ 
$\xi(s)\xi(s) \to \int_{p,q} \xi(p) \xi(q) \exp(i s\cdot(p+q))$, 
carry out $\int_s (...)$, and expand then the resulting
Fourier transform of $g(s-z;M_{uc}^2)$ in $(p+q)^2/M_{uc}^2$. 
Introducing in the end again configuration space, 
\eq\nr{mat1} can be written as 
\be
 \langle {\cal O}(z) \rangle_\4 = \xi(z) \xi(z) \int_s g(s;M_{uc}^2) + 
      \fr1{2d} \partial_\mu^2 [\xi(z)\xi(z)] \int_s s^2 g(s;M_{uc}^2) 
      + ... \;,   \la{mat4}
\ee
with an infinite series of further ``total derivative'' operators.  
To rewrite the total derivative operators which do contribute
in~\eq\nr{match_su3} in a more familiar form, we can use the
equations of motion for the light fields $\xi$, 
$(\partial_\mu^2 -2 M_u^2 )\xi (z) = 0$, given that only 
tree-level light propagators appear in these graphs. 
\eq\nr{mat4} then finally becomes 
\ba
 \langle {\cal O}(z) \rangle_\4& = &  \xi(z) \xi(z) \int_s g(s;M_{uc}^2) \nn
 & + & \Bigl[ \partial_\mu \xi(z)\partial_\mu \xi(z)
      + 2 M_u^2 \xi(z) \xi(z) \Bigr] \fr1{d}  \int_s s^2 g(s;M_{uc}^2) 
      + ...\, .   \la{mat5}
\ea

It may be worth stressing here 
that the total derivative operators in the series of~\eq\nr{mat4} 
{\em do contribute} to correlators of the 
type $\langle {\cal J}^a_\mu(x) \; 
\langle {\cal O}(z) \rangle_\4 \;{\cal J}^b_\nu(y)\rangle$, 
for a fixed $z$. These contributions
are just suppressed by successively 
higher powers of $M_u^2/M_{c}^2$, 
where we count $\partial_\mu \sim M_u$, 
as in the usual SU(3) chiral counting. 


Applying these manipulations to~\eq\nr{expl}, we obtain to the working
order in the chiral expansion (in fact all terms can be assigned
an overall ${\cal O}(M_c^2/(4\pi F)^2)$) and to the relevant
order in the $1/m_c$-expansion 
the result
\ba
 & & \!\!\!\!\! 
 \langle [{\cal O}_{w}]_{\ga 4 4 \gd} \rangle_\4 = 
 \langle [{\cal O}_{w}]_{4 \ga \gd 4} \rangle_\4 = \nn 
 & & \!\! - \fr16 
 \xi_{\gd \gk} 
 \xi_{\gk \ga} 
 \Bigl( 1 + M_u^2 \frac{{\rm d}}{{\rm d} M_c^2} \Bigr)
 \biggl( 
 \partial_\omega^2 \, G(0;M_{c}^2) + 
 M_{c}^2 \int_s [\partial_\omega G(s;M_{c}^2)]^2 + 
  \int_s [\partial_\omega \partial_\psi G(s;M_{c}^2)]^2
 \biggr)  \nn
 & & \!\!
 - 
 \xi_{\gd \gk} 
 \xi_{\gk \ga} 
 \biggl( M_u^2 \int_s [\partial_\omega G(s;M_{c}^2)]^2 \biggr)
 - \fr12  
 \partial_\mu \xi_{\gd \gk} 
 \partial_\mu \xi_{\gk \ga} 
 \, 
 \biggl( 
 G(0;M_{c}^2)+ 
 \int_s [\partial_\omega G(s;M_{c}^2)]^2   
 \biggr)
 \;.  \la{6th} 
\ea
Any other index choice including ``$\4$''
leads to contributions ${\cal O}(M_c^4/(4\pi F_{})^4)$,
which are of higher order in the chiral expansion, while further
orders of $M_u^2/M_c^2$ in the expansion of \eq\nr{expl}
lead to  SU(3) operators which are 
higher-dimensional than those in~\eqs\nr{su3_1}, \nr{su3_3}, 
and thus suppressed by $1/M_c^2$. 

The integrals appearing in~\eq\nr{6th} are all related, 
and can be expressed in 
terms of two basic ones: $G(0;M_{c}^2)$ defined in~\eq\nr{Gx}, 
as well as its derivative with respect to $-M_{c}^2$, 
\be
 B(M_{c}^2) = \int \! \frac{{\rm d}^{d} p}{(2\pi)^{d}}
 \frac{1}{(p^2 + M_{c}^2)^2}. 
\ee
In dimensional regularisation in $d=4-2\epsilon$ dimensions, 
introducing $\bmu^2 = 4\pi e^{-\gamma_E} \mu^2$, 
\ba
 G(0;M_{c}^2) & = & - M_{c}^2 \, \frac{\mu^{-2\epsilon}}{(4\pi)^2} 
 \biggl(\frac{1}{\epsilon} + \ln\frac{\bmu^2}{M_{c}^2} +1 \biggr), \\
 B(M_{c}^2) & = & \frac{\mu^{-2\epsilon} }{(4\pi)^2} 
 \biggl(\frac{1}{\epsilon} + \ln\frac{\bmu^2}{M_{c}^2} \biggr),
\ea
and the relations needed read
\ba
 \partial_\omega^2 G(0;M_{c}^2) & = & M_{c}^2 G(0;M_{c}^2), \\
 \int_s [\partial_\omega G(s;M_{c}^2)]^2  & = & 
 G(0;M_{c}^2) - M_{c}^2 B(M_{c}^2), \\
 \int_s [\partial_\omega \partial_\psi G(s;M_{c}^2)]^2 & = & 
 - 2 M_{c}^2 G(0;M_{c}^2) + M_{c}^4 B(M_{c}^2). 
\ea

Summing now everything together, it is easy to identify the structures 
$\bar {\cal O}_{w}$ and $\bar {\cal O}_{m}$ of \eqs\nr{su3_1}, \nr{su3_3}.
The terms $\sim \xi_{\gd\gk} \xi_{\gk\ga}$ 
are of the form $\bar {\cal O}_{m}$ while 
the terms $\sim  \partial_\mu \xi_{\gd \gk}
 \partial_\mu \xi_{\gk \ga}$ are of the form $\bar {\cal O}_{w}$. 
It is straightforward
to see that the terms on the first row of~\eq\nr{6th} cancel, 
which could have been 
expected, since they do not individually
vanish for $m_u=0$, while $\bar {\cal O}_{m}$ does.
The first term of the second row of \eq\nr{6th} is then 
proportional to $\bar {\cal O}_{m}$, while the last term 
contributes in~\eq\nr{match_su3} just like an SU(3) weak operator 
with the index structure $[\bar {\cal O}_{w}]_{\ga \gk \gk \gd}$.  

Once we also include the SU(4) operators with the index structure 
$[{\cal O}_{w}]_{\ga \gb \gc \gd}$, and 
take into account the redefinition of $F_{}$ from~\eq\nr{barF}, 
we obtain the following matching of the SU(4) and SU(3) operators:
\ba
 [{\cal O}_{w}]_{\a\b\c\d} & \to &  c_1 [\bar{\cal O}_{w}]_{\ga\gb\gc\gd} 
 + c_2 \Bigl( \delta_{\a \4}\delta_{\d \4} 
 [\bar{\cal O}_{w}]_{\gb\gk\gk\gc} + 
 \delta_{\b \4}\delta_{\c \4} 
 [\bar{\cal O}_{w}]_{\ga\gk\gk\gd}
 \Bigr) 
 \nn & & 
 + d_2 \Bigl( \delta_{\a \4}\delta_{\d \4} 
 {[ \bar {\cal O}_{m} ]}_{\gb\gc} + 
 \delta_{\b \4}\delta_{\c \4} 
 {[ \bar {\cal O}_{m} ]}_{\ga\gd}
 \Bigr) , 
 \la{replace}
\ea
where  
\ba
 c_1 & = & 1,  \la{prec1} \\
 c_2 & = & \frac{1}{2 F_{}^2} \Bigl[2 G(0;M_c^2) - M_c^2  B(M_c^2) \Bigr], 
 \la{prec2} \\
 d_2 & = & \frac{1}{2 F_{}^2} \Bigl[ G(0;M_c^2) - M_c^2  B(M_c^2) \Bigr]. 
 \la{pred2}
\ea
In \eq\nr{replace}, we
have left out contributions of the same order 
to the SU(3) singlet operator ${[ \bar {\cal O}_{m} ]}_{\gk\gk}$, 
which can arise for SU(4) 
indices such that $\delta_{\a\c} \delta_{\b\d}  \neq 0$, 
since this situation does not emerge in the realistic strangeness
violating case.

It is worth stressing that all 
order ${\cal O}(M_c^2/(4\pi F_{})^2)$ 
effects have cancelled in~$c_1$, 
after the redefinition of $F_{}^2$. This may seem somewhat miraculous 
since, as far as we can see, the action in~\eq\nr{Lw_XPT} can
in principle also include higher dimensional weak operators such as 
\be
 {\cal H}_w = 
 \frac{K}{F_{}^2} [{\cal O}_{w}]_{\ga\gb\gc\gd} \,
 \tr [\chi^\dagger U + U^\dagger \chi], 
 \la{higher_order_w}
\ee
where $K$ is some dimensionless numerical coefficient. This would 
contribute to $c_1$ at 
the same order in the chiral expansion as we have been 
considering, modifying it to 
\be
 c_1 = 1 + 4 K \frac{M_c^2}{F_{}^2}. \la{c1}
\ee
Thus, the relation in~\eq\nr{prec1} could in principle still 
have a correction of relative order $M_c^2/(4 \pi F_{})^2$, 
but without a logarithm involving $M_c$.

In fact, 
a related ``miracle'' is observed if we consider the correlator
in~\eq\nr{match_su3} with ${\cal H}_w(z) \to [{\cal O}_{w}]_{\a\b\c\d}(z)$,
at the next-to-leading order in the chiral expansion, but
in the case of a degenerate quark mass matrix. Then the part of 
the result which has the same index structure as at tree-level, 
namely $T^a_{\c\a} T^b_{\d\b} + T^a_{\d\b} T^b_{\c\a}$,
is observed to factorise 
into a product of two-point functions of the type in~\eq\nr{bm1}. This 
corresponds to the fact that all 1-loop effects for these
index structures can be accounted for by a redefinition of $F$.
On the contrary, no factorisation takes place for the other flavour 
structures, among them those corresponding to penguin-type charm  
contractions on the side of full QCD.

The coefficients $c_2,d_2$ do
instead contain a logarithm. 
After the inclusion of the proper higher dimensional
operators analogous to~\eq\nr{higher_order_w},
cancelling the ultraviolet divergences, they can be written in the forms 
\be
 c_2 =  - 3 \frac{M_c^2}{(4 \pi F_{})^2} \ln\frac{\Lambda_\chi}{M_c} 
 \;, 
 \qquad
 d_2 = - 2 \frac{M_c^2}{(4 \pi F_{})^2} \ln\frac{\Lambda_\chi}{M_c}
 \;,
 \la{c2}
\ee
where $\Lambda_\chi$ is some (physical) scale. Note 
that although in general 
the $\Lambda_\chi$'s appearing in $c_2,d_2$ are not the same,
we have not introduced two different scales here,  
since we are really only addressing the coefficients of the logarithms, 
and thus 
ignore the unknown finite ``constant'' corrections which differ in each case. 

Let us finally remark that 
a weak operator ${\cal O}_m$, of the form 
in~\eq\nr{O2_XPT}, can also contribute to the SU(3) effective 
weak Hamiltonian $\bar {\cal H}_w$. 
For the mass matrix in~\eq\nr{Mass} and with our power-counting
convention, there is indeed a tree-level effect of the type
\ba
 \langle [{\cal O}_m]_{\a\b\c\d} \rangle_\4 & = & - \frac{m_c^2}{F_{}^2}
  \Bigl( \delta_{\a \4}\delta_{\d \4} 
  {[ \bar {\cal O}_{m} ]}_{\gb\gc} + 
  \delta_{\b \4}\delta_{\c \4} 
  {[ \bar {\cal O}_{m} ]}_{\ga\gd}
  \Bigr)
 \;. \la{o2contr}
\ea
In the region of small $m_c$ that we are working this is, however, 
parametrically smaller than the contribution with the same 
structure in~\eq\nr{replace}, with the coefficient $d_2$ 
in~\eq\nr{pred2}, which is ${\cal O}(m_c)$ rather than ${\cal O}(m_c^2)$.
Therefore, we can neglect~\eq\nr{o2contr}.

\subsection{SU(3) classification of the effective theory}
\la{su3_class}

We now continue with the full SU(4) weak Lagrangian
from~\eq\nr{Lw_XPT}, written in the form 
\be
 {\cal H}_w
 =  \sum_{\sigma = \pm 1} g_{w}^\sigma c_{\a\b\c\d}
 (P_2^\sigma P_1^\sigma)_{\a\b\c\d;\ta\tb\tc\td}\;
 {[ {\cal O}_{w} ]}_{\ta\tb\tc\td}, 
 \la{start_su3}
\ee
where $P_1^\sigma,P_2^\sigma$ are defined in~\eqs\nr{P1}, \nr{P2}.
{}From the previous section we know that to  
order ${\cal O}(m_c)$, we can  
summarise the result of matching as the replacement
in~\eq\nr{replace},
when we move from ${\cal H}_w$ to $\bar {\cal H}_w$,
with $c_1,c_2,d_2$ from \eqs\nr{c1}, \nr{c2}.
We thus insert \eq\nr{replace} into~\eq\nr{start_su3}, and carry 
then out a decomposition into irreducible representations of 
SU(3) according to \ref{app:su3}. The result is
\ba
 && \hspace*{-2cm} \sum_{\sigma = \pm 1} g_{w}^\sigma 
 (P_2^\sigma P_1^\sigma)_{\a\b\c\d;\ta\tb\tc\td}\;
 {[ {\cal O}_{w} ]}_{\ta\tb\tc\td} \to \nn
 && 
 + c_1 \, g_{w}^+ [\; \hat {\!\bar {\cal O}}_{w}]^+_{\ga\gb\gc\gd} \nn 
 && 
 + \frac{c_1 - 5c_2}{30} g_{w}^+ \biggl[ \Bigl( 
 \delta_{\ga\gc} {[ \bar {\cal R}_{w}]} ^+_{\gb\gd} + 
 \delta_{\gb\gd} {[ \bar {\cal R}_{w}]}^+_{\ga\gc} + 
 \delta_{\ga\gd} {[ \bar {\cal R}_{w}]}^+_{\gb\gc} + 
 \delta_{\gb\gc} {[ \bar {\cal R}_{w}]}^+_{\ga\gd}
 \Bigr) \nn 
 && \hspace*{1.6cm} - 5  \Bigl( 
 \delta_{\a \4} \delta_{\c \4} {[ \bar {\cal R}_{w}]}^+_{\gb\gd} + 
 \delta_{\b \4} \delta_{\d \4} {[ \bar {\cal R}_{w}]}^+_{\ga\gc} + 
 \delta_{\a \4} \delta_{\d \4} {[ \bar {\cal R}_{w}]}^+_{\gb\gc} + 
 \delta_{\b \4} \delta_{\c \4} {[ \bar {\cal R}_{w}]}^+_{\ga\gd}
 \Bigr) \biggr] \nn 
 && + \frac{c_1 - c_2}{2} g_{w}^- \biggl[ \Bigl( 
 \delta_{\ga\gc} {[ \bar {\cal R}_{w}]}^-_{\gb\gd} + 
 \delta_{\gb\gd} {[ \bar {\cal R}_{w}]}^-_{\ga\gc} - 
 \delta_{\ga\gd} {[ \bar {\cal R}_{w}]}^-_{\gb\gc} - 
 \delta_{\gb\gc} {[ \bar {\cal R}_{w}]}^-_{\ga\gd}
 \Bigr) \nn 
 && \hspace*{1.6cm} - \Bigl( 
 \delta_{\a \4} \delta_{\c \4} {[ \bar {\cal R}_{w}]}^-_{\gb\gd} + 
 \delta_{\b \4} \delta_{\d \4} {[ \bar {\cal R}_{w}]}^-_{\ga\gc} - 
 \delta_{\a \4} \delta_{\d \4} {[ \bar {\cal R}_{w}]}^-_{\gb\gc} - 
 \delta_{\b \4} \delta_{\c \4} {[ \bar {\cal R}_{w}]}^-_{\ga\gd}
 \Bigr) \biggr] \nn 
 && 
 - \frac{d_2}{12} g_{w}^+ \biggl[ \Bigl( 
 \delta_{\ga\gc} {[\; \hat{\!\bar {\cal O}}_{m}]}_{\gb\gd} + 
 \delta_{\gb\gd} {[\; \hat{\!\bar {\cal O}}_{m}]}_{\ga\gc} + 
 \delta_{\ga\gd} {[\; \hat{\!\bar {\cal O}}_{m}]}_{\gb\gc} + 
 \delta_{\gb\gc} {[\; \hat{\!\bar {\cal O}}_{m}]}_{\ga\gd}
 \Bigr) \nn 
 && \hspace*{0.8cm} - 5  \Bigl( 
 \delta_{\a \4} \delta_{\c \4} {[\; \hat{\!\bar {\cal O}}_{m}]}_{\gb\gd} + 
 \delta_{\b \4} \delta_{\d \4} {[\; \hat{\!\bar {\cal O}}_{m}]}_{\ga\gc} + 
 \delta_{\a \4} \delta_{\d \4} {[\; \hat{\!\bar {\cal O}}_{m}]}_{\gb\gc} + 
 \delta_{\b \4} \delta_{\c \4} {[\; \hat{\!\bar {\cal O}}_{m}]}_{\ga\gd}
 \Bigr) \biggr] \nn 
 && + \frac{d_2}{4} g_{w}^- \biggl[ \Bigl( 
 \delta_{\ga\gc} {[\; \hat{\!\bar {\cal O}}_{m}]}_{\gb\gd} + 
 \delta_{\gb\gd} {[\; \hat{\!\bar {\cal O}}_{m}]}_{\ga\gc} - 
 \delta_{\ga\gd} {[\; \hat{\!\bar {\cal O}}_{m}]}_{\gb\gc} - 
 \delta_{\gb\gc} {[\; \hat{\!\bar {\cal O}}_{m}]}_{\ga\gd}
 \Bigr) \nn 
 && \hspace*{1.0cm} - \Bigl( 
 \delta_{\a \4} \delta_{\c \4} {[\; \hat{\!\bar {\cal O}}_{m}]}_{\gb\gd} + 
 \delta_{\b \4} \delta_{\d \4} {[\; \hat{\!\bar {\cal O}}_{m}]}_{\ga\gc} - 
 \delta_{\a \4} \delta_{\d \4} {[\; \hat{\!\bar {\cal O}}_{m}]}_{\gb\gc} - 
 \delta_{\b \4} \delta_{\c \4} {[\; \hat{\!\bar {\cal O}}_{m}]}_{\ga\gd}
 \Bigr) \biggr] 
 ,\la{full_res} 
\ea
where $\hat {\!\bar {\cal O}}_{w}$ and $\bar {\cal R}_{w}$
are defined in \ref{app:su3}, and 
\be
 {[\, \hat{\!\bar {\cal O}}_{m}]}_{\ga\gc} \equiv 
 {[ \bar {\cal O}_{m}]}_{\ga\gc} - \fr13 \delta_{\ga\gc}
 {[ \bar {\cal O}_{m}]}_{\gk\gk}
 \;. \la{su3_3_trls}
\ee
For brevity, we have left out all SU(3) singlet 
operators, $g_{w}^+ {[ \bar {\cal S}_{w}]}^+, 
g_{w}^+ {[ \bar {\cal O}_{m}]}_{\gk\gk}$, since their coefficients
vanish if $\delta_{\a\c}\delta_{\b\d} = \delta_{\a\d}\delta_{\b\c} = 0$, 
as is the case for the $c_{\a\b\c\d}$ relevant for physical
strangeness violating interactions.  

\section{Implications for the $\Delta I = 1/2$ rule} 
\la{se:physics}

Let us finally consider the physical 
choice of indices, according to~\eq\nr{physind}. 
Inserting~\eq\nr{physind}  into \eq\nr{full_res} and
employing~\eq\nr{traceless_su3}, we obtain 
\be
 \bar {\cal H}_w = 
 c_1\, g_{w}^+ [\; \hat {\!\bar {\cal O}}_{w}]^+_{\3\2\2\1} + 
 \biggl( \frac{c_1 - 5 c_2}{5} g_{w}^+ + ({c_1-c_2}) g_{w}^- \biggr)
 \, {[ \bar {\cal R}_{w} ]}_{\3\1}^+ - 
 \frac{d_2}{2} (g_{w}^+ + g_{w}^-)  
 {[\, \hat{\!\bar {\cal O}}_{m}]}_{\3\1} \;.  \la{result0}
\ee
Here and in the remainder, we leave out the
Hermitean conjugated part of the weak Hamiltonian.
The first operator in~\eq\nr{result0}
is defined as in~\eqs\nr{OP1}--\nr{P2} with $\Nf = 3$
(and transforms under the representation 27), while the second is 
defined as in \eq\nr{cRac} (and transforms under the representation 8): 
\ba
 {[\; \hat {\!\bar {\cal O}}_{w}]}^+_{\3\2\2\1} & = &  \fr12 \Bigl( 
 {[ \bar {\cal O}_{w} ]}_{\3\2\2\1} + 
 {[ \bar {\cal O}_{w} ]}_{\3\2\1\2} - \fr15 
 {[ \bar {\cal O}_{w} ]}_{\3\gk\gk\1} 
 \Bigr) = 
 \fr35\Bigl(  
 {[ \bar {\cal O}_{w} ]}_{\3\2\1\2} + 
 \fr23 {[ \bar {\cal O}_{w} ]}_{\3\2\2\1}
 \Bigr)
 \;, \la{formofO} \\   
 {[ \bar {\cal R}_{w} ]}_{\3\1}^+ & = & \fr12 
 {[ \bar {\cal O}_{w} ]}_{\3\gk\gk\1}
 \;, 
\ea
where $\bar {\cal O}_{w}$ is from~\eq\nr{su3_1}.
In \eq\nr{formofO} we have displayed two different forms 
sometimes appearing in the literature, equivalent 
on account of \eq\nr{tless_su3}. 
The third operator in~\eq\nr{result0}
is defined in \eqs\nr{su3_3}, \nr{su3_3_trls}, 
and also transforms under the representation 8.
The three operators in \eq\nr{result0} are directly proportional to the 
standard ones defined, e.g., in Ref.~\cite{b}.

Inserting the values of $c_1,c_2,d_2$ from \eqs\nr{c1}, \nr{c2}
and keeping only the ``chiral logarithms'',  
we finally obtain
\ba
 \bar {\cal H}_w & = & 
 g_{w}^+ \, {[\; \hat {\!\bar {\cal O}}_{w}]}^+_{\3\2\2\1}\nn 
 & + &  
 \biggl[ 
 \fr15 {g_{w}^+} \Bigl(1 + {15} \frac{M_c^2}{(4\pi F_{})^2}
                 \ln\frac{\Lambda_\chi}{M_c} \Bigr) + 
 {g_{w}^-} \Bigl( 1 + {3} \frac{M_c^2}{(4\pi F_{})^2}
                 \ln\frac{\Lambda_\chi}{M_c} \Bigr) 
 \biggr]
 {[ \bar {\cal R}_{w} ]}_{\3\1}^+ \nn 
 & + & 
 \biggl[
 (g_{w}^+ + g_{w}^-) \Bigl( \frac{M_c^2}{(4\pi F_{})^2} 
 \ln\frac{\Lambda_\chi}{M_c} \Bigr) \biggr]
 {[\, \hat{\!\bar {\cal O}}_{m}]}_{\3\1} .  
 \la{final_result}
\ea 
Carrying out furthermore an isospin decomposition  (cf.~\ref{app:su2}),
the last two operators obviously 
mediate only transitions with $\Delta I = 1/2$, 
but the first one has parts both with $\Delta I = 1/2$
and $\Delta I = 3/2$. On account of~\eq\nr{tless_su3}
it can, however, be written as 
\ba
 {[\; \hat {\!\bar {\cal O}}_{w}]}^+_{\3\2\2\1} & = & 
 \frac{1}{3} 
 \Bigl( 
  2 {[ \;  \hat {\!\bar {\cal O}}_{w}]}_{\3\2\2\1}^+ - 
 {[ \; \hat {\!\bar {\cal O}}_{w}]}_{\3\1\1\1}^+ 
 \Bigr)_{\Delta I = 3/2} \nn 
 & + &  
 \frac{1}{15} 
 \Bigl( 
 2 {[ \;\hat {\!\bar {\cal O}}_{w}]}_{\3\2\2\1}^+ + 
 2 {[ \;\hat {\!\bar {\cal O}}_{w}]}_{\3\1\1\1}^+ - 
 3 {[ \;\hat {\!\bar {\cal O}}_{w}]}_{\3\3\3\1}^+
 \Bigr)_{\Delta I = 1/2}. 
\ea
Using~\eq\nr{i32}, we can see that (as indicated by the notation) 
the first row is purely $\Delta I = 3/2$, while the second one is purely 
$\Delta I = 1/2$, completing the isospin decomposition.

\vspace*{0.5cm}

We try finally to make contact with the standard notation in literature. 
Let us rewrite \eq\nr{final_result} 
(apart from ${[\, \hat{\!\bar {\cal O}}_{m}]}_{\3\1}$, which does not 
contribute to physical kaon decays~\cite{wilson,rc})
in the form 
\be
  \bar {\cal H}_w \equiv  2 \sqrt{2} G_F V_\rmi{ud} V^*_\rmi{us}
  \biggl\{ 
  2 g_8 {[ \bar {\cal R}_{w} ]}_{\3\1}^+ + 
  \fr53 g_{27} [\; \hat {\!\bar {\cal O}}_{w}]^+_{\3\2\2\1}
  \biggr\} \;,
\ee
where $g_{8},g_{27}$ are dimensionless coefficients, 
normalised according to their standard definitions~\cite{lo,1loop}.
In addition we define 
\be
 g_{w}^\pm \equiv 2 \sqrt{2} G_F V_\rmi{ud} V^*_\rmi{us}\, \hat g_{w}^\pm \;,
\ee
so that the tree-level values (or $\alpha_s\sim 1/\Nc\to 0$ limits) 
are $\hat g_{w}^\pm = 1$. \eq\nr{final_result} then implies
\ba
 g_8 & = & \fr12 
 \biggl[ 
 \fr15 {\hat g_{w}^+} \Bigl(1 + {15} \frac{M_c^2}{(4\pi F_{})^2}
                 \ln\frac{\Lambda_\chi}{M_c} \Bigr) + 
 {\hat g_{w}^-} \Bigl( 1 + {3} \frac{M_c^2}{(4\pi F_{})^2}
                 \ln\frac{\Lambda_\chi}{M_c} \Bigr) 
 \biggr] 
 \;, \la{relg8}\\ 
 g_{27} & = & \fr35 \hat g_{w}^+ 
 \;. \la{relg27} 
\ea 

Now, $g_8$ should phenomenologically be much 
larger in absolute value than $g_{27}$. This can perhaps most
easily be understood by writing $\xi$ in the meson basis and 
expanding the operators to the third order, whereby it is easy 
to verify that the very slow $\Delta I = 3/2$ decay $K^+\to\pi^0\pi^+$ is 
directly proportional to $g_{27}$, while the much faster
$\Delta I =1/2$ decays of
$K^0_S$ get a comparable contribution both from $g_8$ and $g_{27}$. 
More quantitatively, a leading order analysis~\cite{lo}, supplemented 
by phenomenologically determined large phase factors~\cite{gm} in the 
amplitudes, suggests the well-known values
\ba
 |g_8|    & \approx & 5.1 \;, \\
 |g_{27}| & \approx & 0.29 \;.
\ea
It has been argued that 1-loop corrections in 
chiral perturbation theory are large~\cite{p4,1loop,pp}, 
and one can therefore get agreement
with experimental data on partial decay widths even with somewhat
less differing values of $g_8$ and $g_{27}$, 
but a sizeable hierarchy still remains.

\eqs\nr{relg8}, \nr{relg27} now indicate that the charm quark mass
can contribute to this hierarchy. Indeed, 
we observe that 
even if the SU(4) values were degenerate, 
$\hat g_{w}^+ \approx \hat g_{w}^-$, there is a logarithmically 
enhanced linear term in $g_8$, but none in $g_{27}$.
Inserting $\Sigma \approx (250 \mbox{ MeV})^3$, 
$F_{}\approx 93$~MeV, we note that 
${M_c^2}/{(4\pi F_{})^2}\sim m_c/(760 \mbox{ MeV})$, which 
means that the correction factors in \eq\nr{relg8}
are rather substantial, as soon as $m_c$ equals a few hundred MeV. 
The situation is illustrated numerically in~\fig\ref{fig:R}. 

\begin{figure}[t]

\begin{center}
\epsfig{file=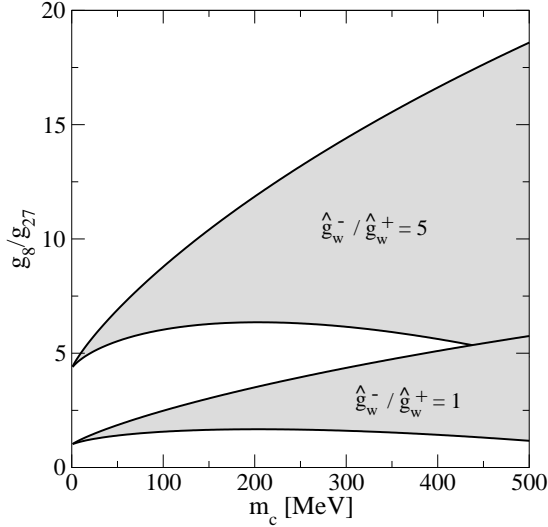,height=7cm}%
\end{center}

\caption[a]{\small
The ratio $g_8/g_{27}$ of SU(3) weak couplings in the chiral limit,
for two different values of the ratio $\hat g_{w}^- / \hat g_{w}^+$
of SU(4) weak couplings in the chiral limit. We have assumed  
$\Sigma \approx (250 \mbox{ MeV})^3$, $F_{}\approx 93$~MeV, and varied
the scale $\Lambda_\chi$, accounting for unknown higher order 
low-energy couplings, in the range from $\Lambda_\chi = 1$~GeV (lower edges 
of the bands) to $\Lambda_\chi = 4$~GeV (upper edges of the bands).
}

\la{fig:R}
\end{figure}

\section{Conclusions}
\la{se:concl}

It is generally believed~\cite{itep,hg,revs} 
that the charm quark plays an important 
role in the $\Delta I = 1/2$ rule, observed in non-leptonic
weak kaon decays, $K\to \pi\pi$. The purpose of this short note 
has been to investigate one aspect of this role. The 
philosophy~\cite{lattice} has been to assume
that the weak chiral Hamiltonian is known in the unphysical 
SU(4) limit of a very light charm quark. We then start
to increase the charm mass $m_c$, taking it eventually to be much larger
than $m_u, m_d$ and $m_s$, but still small 
enough such that chiral perturbation theory is applicable. 
In this limit, the charm quark can be integrated 
out within an SU(4) chiral effective theory, to 
obtain an SU(3) chiral effective theory.  

We have found that the charm quark
does have a distinctive effect on the 
weak interaction Hamiltonian of this SU(3) theory. The $\Delta I=1/2$ part 
obtains a logarithmically enhanced contribution, ${\cal O}(m_c\ln(1/m_c))$, 
while there are no logarithms in the $\Delta I=3/2$ part, 
which is at most $\mathcal{O}(m_c)$
(\eqs\nr{relg8}, \nr{relg27}). Thus, the ratio of $\Delta I=1/2$ and
$\Delta I=3/2$ amplitudes departs from the SU(4) limit as 
$\sim m_c\ln(1/m_c)$. Moreover, the 
numerical coefficients of the logarithmic contributions 
are fairly large. The effect comes from penguin-type graphs, 
the last three in~\fig\ref{fig:effc}, as suggested by weak coupling
considerations a long time ago~\cite{itep}.

The actual values of the 1-loop corrections
are, of course, not computable within our framework, because the unknown 
higher order couplings of the SU(4) chiral Lagrangian enter
at the same order. In addition, once the charm quark mass 
is increased to its physical value, chiral perturbation theory breaks 
down. Therefore, to really settle from first principles the issue
of how important the charm quark is, requires a 
lattice study~\cite{lattice,as}. 
We may note, however, that since Ref.~\cite{itep} relies on the 
weak coupling expansion and is thus trustworthy for $m_c$ larger than its 
physical value, while our approach relies on the chiral expansion and 
is trustworthy for $m_c$ smaller than its physical value, the
importance of $m_c$ is at least confirmed in two complementary limits.
Further evidence for its importance comes from the large-$\Nc$ 
approach (\cite{bbg,hpr} and references therein), attempting 
to interpolate between these limits.

To summarise, the results obtained here indicate
that together with all the other effects, the dynamics 
related to the charm quark may play a role in the enhancement
observed in $\Delta I = 1/2$ decays, providing thus
additional motivation for lattice studies. 

\section*{Acknowledgements}

This work originated as a part of a bigger effort  whose goal
is to extract low-energy parameters of QCD from numerical 
simulations with so-called Ginsparg-Wilson fermions. 
We are indebted to our collaborators on this,  
L.~Giusti, P.~Weisz and H.~Wittig, for useful comments. 
The basic ideas of the general approach were developed 
in collaboration with M. L\"uscher; we would like to 
thank him for his input and for many valuable suggestions. 
We are also grateful to L.~Lellouch for discussions. 
P.~H.\ was supported in part by the CICYT (Project No.\ FPA2002-00612) and 
by the Generalitat Valenciana (Project No.\ CTIDIA/2002/5). 


\appendix
\renewcommand{\thesection}{Appendix~\Alph{section}}
\renewcommand{\thesubsection}{\Alph{section}.\arabic{subsection}}
\renewcommand{\theequation}{\Alph{section}.\arabic{equation}}


\newpage

\section{SU(4) classification}
\la{app:su4}

We reiterate in this Appendix some essential aspects of 
the SU(4) classification of four quark operators.
We follow the tensor method discussed, e.g., in Ref.~\cite{hg2}. 
We also greatly profited from the presentation in Ref.~\cite{ml}.

We start from an operator $O_{\a\b\c\d}$, assumed
symmetric under $(\a\leftrightarrow\b,\c\leftrightarrow\d)$, 
and transforming under 
${\bf 4^*} \otimes {\bf 4^*} \otimes {\bf 4} \otimes {\bf 4}$
of SU(4). We then define the projected operators 
\ba
 {O}^\sigma_{\a\b\c\d} & \equiv &  
 (P_1^\sigma)_{\a\b\c\d;\ta\tb\tc\td}\;
 {O}_{\ta\tb\tc\td}, \la{O_sigma} \la{OP1} \\
 \hat {O}^\sigma_{\a\b\c\d} & \equiv &  
 (P_2^\sigma)_{\a\b\c\d;\ta\tb\tc\td}\;
 {O}^\sigma_{\ta\tb\tc\td},  \la{O_project} \la{OP2}
\ea
where $\sigma=\pm 1$. Here, 
with some redundancy in the symmetries of $P_1^\sigma$, 
\ba
 & & (P_1^\sigma)_{\a\b\c\d ; \ta\tb\tc\td} \equiv 
 \fr14 (\delta_{\a\ta} \delta_{\b\tb} + 
 \sigma \delta_{\a\tb}\delta_{\b\ta})
 (\delta_{\c\tc} \delta_{\d\td} + 
 \sigma \delta_{\c\td} \delta_{\d\tc}), \la{P1} \\
 & &  (P_2^\sigma)_{\a\b\c\d;\ta\tb\tc\td} \equiv  
 \delta_{\a\ta} \delta_{\b\tb} \delta_{\c\tc} \delta_{\d\td} 
 + \frac{1}{(\Nf + 2\sigma)(\Nf + \sigma)}
 ( 
 \delta_{\a\c} \delta_{\b\d} + \sigma
 \delta_{\a\d} \delta_{\b\c}) \delta_{\ta\tc} \delta_{\tb\td}
 \nn
 & & ~~~~~~~~ - \frac{1}{\Nf + 2\sigma}
 (
 \delta_{\a\c} \delta_{\b\tb}  \delta_{\d\td} \delta_{\ta\tc}+
 \delta_{\b\d} \delta_{\a\ta} \delta_{\c\tc} \delta_{\tb\td}+
 \sigma \delta_{\a\d} \delta_{\b\tb} \delta_{\c\td} \delta_{\ta\tc}+
 \sigma \delta_{\b\c} \delta_{\a\ta} \delta_{\d\tc} \delta_{\tb\td}
 ), \hspace*{1.0cm} \la{P2}
\ea
where $\Nf=4$. Let us also denote
\ba
 S^\sigma & = & O^\sigma_{\k\l\k\l}, \\
 R^\sigma_{\a\c} & = & O^\sigma_{\a\k\c\k} - \frac{1}{\Nf} 
 \delta_{\a\c} S^\sigma. 
\ea
Then we can decompose any operator
$O_{\a\b\c\d}$ into irreducible representations, as
\ba
 O_{\a\b\c\d} & = & \sum_{\sigma = \pm 1}\biggl[ 
 \hat O^\sigma_{\a\b\c\d} + \frac{1}{\Nf(\Nf+\sigma)}
 (\delta_{\a\c}\delta_{\b\d} + \sigma \delta_{\a\d}\delta_{\b\c}) S^\sigma \nn 
  & &  \hspace*{2cm} + \frac{1}{\Nf + 2\sigma} 
  (\delta_{\a\c} R^\sigma_{\b\d} + \delta_{\b\d} R^\sigma_{\a\c} + 
  \sigma \delta_{\a\d} R^\sigma_{\b\c} + \sigma \delta_{\b\c} R^\sigma_{\a\d})
 \biggr].
\ea
Here the representation $\hat O^+_{\a\b\c\d}$ has dimension 84, 
$\hat O^-_{\a\b\c\d}$ 20, 
$R^\pm_{\b\d}$'s 15, 
and $S^\pm$ are singlets.

For a sum over irreducible representations 
with some weights, $c_{\a\b\c\d} \hat O^\sigma_{\a\b\c\d}$, 
we can take advantage of 
the fact $P_2^\sigma P_1^\sigma$ is a projection operator 
to symmetrise the coefficients: 
\be
 c_{\a\b\c\d} \hat O^\sigma_{\a\b\c\d} = 
 \Bigl[ c_{\a\b\c\d} (P_2^\sigma P_1^\sigma)_{\a\b\c\d;\ta\tb\tc\td} \Bigr] 
 \hat O^\sigma_{\ta\tb\tc\td} \equiv 
 \hat c^\sigma_{\ta\tb\tc\td}
 \hat O^\sigma_{\ta\tb\tc\td} = 
 \hat c^\sigma_{\ta\tb\tc\td}
 O_{\ta\tb\tc\td}. \la{sym_c}
\ee
However, sometimes it may be more convenient not to carry out 
any symmetrisation. 

Finally, we note that in the chiral theory, the matrix
$(\partial_\mu U U^\dagger)_{\a\c}$ is traceless.
For the operator ${[ {\cal O}_{w} ]}_{\a\b\c\d}$
in~\eq\nr{O_XPT}, therefore, 
\be
 {[ {\cal O}_{w} ]}_{\a\k\c\k} = {[ {\cal O}_{w} ]}_{\k\b\k\d} = 0.
\ee
Consequently, if we use ${[ {\cal O}_{w} ]}_{\a\b\c\d}$ in the role 
of $O_{\a\b\c\d}$ above, 
\be
 {[ {\cal O}_{w} ]}^\sigma_{\a\k\c\k} = 
 \frac{\sigma}{2} {[ {\cal O}_{w} ]}_{\a\k\k\c}, \quad
 {[ {\cal S}_{w} ]}^\sigma = \frac{\sigma}{2} {[ {\cal O}_{w} ]}_{\l\k\k\l},
 \quad
 {[ {\cal R}_{w} ]}^+_{\a\c} =  -{[ {\cal R}_{w} ]}^-_{\a\c}.
 \la{traceless}
\ee


\section{SU(3) classification}
\la{app:su3}

Let us then consider the SU(3) classification of
tensors of the form $\bar O_{\ga \gb \gc \gd}$, 
symmetric under $(\ga\leftrightarrow\gb,
\gc\leftrightarrow\gd)$, and transforming under 
${\bf 3^*} \otimes {\bf 3^*} \otimes {\bf 3} \otimes {\bf 3}$ of SU(3). 

The classification of such operators follows almost immediately 
from \ref{app:su4}, by simply replacing $\Nf \to 3$. There is
only one major difference: the antisymmetric tensor 
$\hat {\!\bar O}^-_{\ga \gb \gc \gd} $
vanishes identically. The reason is that 
(as can be understood for instance by contracting 
with  $\epsilon_{\gk\ga \gb} \epsilon_{\gl\gc \gd}$)
it corresponds to a representation with dimension 8 just like  
$\bar R^-_{\gl\gk}$, but all such representations have already
been subtracted by the projection operator in~\eq\nr{P2}. 

Consequently, the general reduction now proceeds as
\ba
 \bar O_{\ga\gb\gc\gd} & = & 
 \hat {\!\bar O}^+_{\ga\gb\gc\gd} +
 \sum_{\sigma = \pm 1}\biggl[ 
 \frac{1}{3(3+\sigma)}
 (\delta_{\ga\gc}\delta_{\gb\gd} + 
  \sigma \delta_{\ga\gd}\delta_{\gb\gc}) \bar S^\sigma \nn 
  & & \hspace*{2cm} + \frac{1}{3 + 2\sigma} 
  (\delta_{\ga\gc} \bar R^\sigma_{\gb\gd} + 
  \delta_{\gb\gd} \bar R^\sigma_{\ga\gc} + 
  \sigma \delta_{\ga\gd} \bar R^\sigma_{\gb\gc} + \sigma \delta_{\gb\gc} 
  \bar R^\sigma_{\ga\gd})
 \biggr],
\ea
where $\;\hat {\!\bar O}^+_{\ga\gb\gc\gd}$ transforms
under the  representation with the dimension 27, and
\ba
 \bar S^\sigma & = & \bar 
 O^\sigma_{\gk\gl\gk\gl}, \\
 \bar R^\sigma_{\ga\gc} & = & 
 \bar O^\sigma_{\ga\gk\gc\gk} - 
 \frac{1}{3} \delta_{\ga\gc} \bar S^\sigma. \la{cRac}
\ea
Here $\bar R^\pm_{\ga\gc}$'s have the dimension 8, 
while $\bar S^\sigma$ are singlets. 

Finally, let us again note that in the chiral theory,
i.e.\ if we replace  
$\bar O_{\ga\gb\gc\gd} \to {[ \bar{\cal O}_{w} ]}_{\ga\gb\gc\gd}$, 
\be
 {[ \bar {\cal O}_{w} ]}_{\ga\gk\gc\gk} = 
 {[ \bar {\cal O}_{w} ]}_{\gk\gb\gk\gd} = 0, \la{tless_su3}
\ee
so that
\be
 {[ \bar {\cal O}_{w} ]}^\sigma_{\ga\gk\gc\gk} = 
 \frac{\sigma}{2} {[ \bar {\cal O}_{w} ]}_{\ga\gk\gk\gc}, \quad
 {[ \bar {\cal S}_{w} ]}^\sigma = \frac{\sigma}{2} 
 {[ \bar {\cal O}_{w} ]}_{\gl\gk\gk\gl}, \quad
 {[ \bar {\cal R}_{w} ]}^+_{\ga\gc} = 
 - {[ \bar {\cal R}_{w} ]}^-_{\ga\gc}.
 \la{traceless_su3}
\ee


\section{SU(2) classification}
\la{app:su2}

We end by considering the SU(2) classification of operators
of the type ${Q}_{\ua\ub\uc}$, $\ua,\ub,\uc = 1,2$, transforming
under ${\bf 2^*} \otimes {\bf 2} \otimes {\bf 2}$. 
We may define 
${Q}_{\ua\ub\uc}^\sigma = (1/2) 
({Q}_{\ua\ub\uc} + \sigma {Q}_{\ua\uc\ub})$.
Then ${Q}_{\ua\ub\uc}$ can be written as 
\be
 {Q}_{\ua\ub\uc} = \hat {Q}_{\ua\ub\uc}^+ + 
 \sum_{\sigma = \pm 1} \frac{1}{2+\sigma} 
 \Bigl( 
 \delta_{\ua\ub}{Q}_{\uk\uk\uc}^\sigma + \sigma
 \delta_{\ua\uc}{Q}_{\uk\uk\ub}^\sigma 
 \Bigr),  \la{i32} 
\ee
where $\hat {Q}_{\ua\ub\uc}^+$ is traceless, 
$\hat {Q}_{\uk\uk\uc}^+ = \hat {Q}_{\uk\ub\uk}^+ = 0$. 
It can be seen to have 4 independent components, and it 
thus isolates the representation with $I = 3/2$.


\end{document}